\documentclass[journal,twoside,web]{ieeetran}
\usepackage{cite}
\usepackage{amsmath,amssymb,amsfonts}
\usepackage{graphicx}
\usepackage{dcolumn}% Align table columns on decimal point
\usepackage{bm}% bold math
\usepackage{amsmath}
\usepackage[T1]{fontenc}
\usepackage{float}
\usepackage{siunitx}
\usepackage{hyperref}
\usepackage{mathabx}

\begin{document}
%\history{Date of publication xxxx 00, 0000, date of current version xxxx 00, 0000.}
%\doi{10.1109/ACCESS.2017.DOI}

\title{Angular harmonic Hall voltage and magnetoresistance measurements of Pt/FeCoB and Pt-Ti/FeCoB bilayers for spin Hall conductivity determination}

%\author{\uppercase{Witold Skowro\'{n}ski}\authorrefmark{1},
%\uppercase{Krzysztof Grochot\authorrefmark{1,2}, Piotr Rzeszut\authorrefmark{1}, Stanisław Łazarski\authorrefmark{1},  Grzegorz Gajoch\authorrefmark{1}, Cezary Worek\authorrefmark{1}, J\"{u}rgen Langer\authorrefmark{3}, Berthold Ocker\authorrefmark{3}, and Mehran Vafaee\authorrefmark{3}}}
%\address[1]{AGH University of Science and Technology, Institute of Electronics, Al. Mickiewicza 30, 30-059 Krak\'{o}w, Poland}
%\address[2]{AGH University of Science and Technology, Faculty of Physics and Applied Computer Science, Al. Mickiewicza 30, 30-059 Krak\'{o}w, Poland}
%\address[3]{Singulus Technologies AG, Hanauer Landstrasse 103, Kahl am Main, 63796, Germany}}
%\tfootnote{This work is supported by the National Science Centre, grant No. UMO-2015/17/D/ST3/00500, Poland. K.G. and S.Ł. acknowledge support from National Science Centre, grant No. UMO-2016/23/B/ST3/01430, Poland. Nanofabrication was performed at the Academic Centre for Materials and Nanotechnology of AGH University of Science and Technology.}

%\markboth
%{W. Skowroński \headeretal: Angular harmonic Hall voltage and magnetoresistance measurements of Pt/FeCoB and Pt-Ti/FeCoB bilayers...}
%{W. Skowroński \headeretal: Angular harmonic Hall voltage and magnetoresistance measurements of Pt/FeCoB and Pt-Ti/FeCoB bilayers...}

%\corresp{Corresponding author: Witold Skowroński (e-mail: skowron@agh.edu.pl).}

\author{Witold Skowro\'{n}ski, Krzysztof Grochot, Piotr Rzeszut, Stanisław Łazarski,  Grzegorz Gajoch, Cezary Worek, Jarosław Kanak, Tomasz Stobiecki, J\"{u}rgen Langer, Berthold Ocker, and Mehran Vafaee \thanks{Manuscript received May 27, 2021 This work is supported by the National Science Centre, grant No. UMO-2015/17/D/ST3/00500, Poland. K.G., S.Ł. and T.S. acknowledge support from National Science Centre, grant No. UMO-2016/23/B/ST3/01430, Poland. Nanofabrication was performed at the Academic Centre for Materials and Nanotechnology of AGH University of Science and Technology. } \thanks{W. Skowroński, K. Grochot, P. Rzeszut, G. Gajoch, C. Worek, J. Kanak and T. Stobiecki are with AGH University of Science and Technology, Institute of Electronics, Al. Mickiewicza 30, 30-059 Krak\'{o}w, Poland (e-mail: skowron@agh.edu.pl) } \thanks{K. Grochot and T. Stobiecki are also with AGH University of Science and Technology, Faculty of Physics and Applied Computer Science, Al. Mickiewicza 30, 30-059 Krak\'{o}w, Poland} \thanks{J. Langer, B. Ocker and M. Vafaee are with Singulus Technologies AG, Hanauer Landstrasse 103, Kahl am Main, 63796, Germany (e-mail: mehran.khanjani@singulus.de)}}
%\thanks{Manuscript received May 27, 2021 This work is supported by the National Science Centre, grant No. UMO-2015/17/D/ST3/00500, Poland. K.G., S.Ł. and T.S. acknowledge support from National Science Centre, grant No. UMO-2016/23/B/ST3/01430, Poland. Nanofabrication was performed at the Academic Centre for Materials and Nanotechnology of AGH University of Science and Technology. }
%\thanks{W. Skowroński, K. Grochot, P. Rzeszut, G. Gajoch, C. Worek, J. Kanak and T. Stobiecki are with AGH University of Science and Technology, Institute of Electronics, Al. Mickiewicza 30, 30-059 Krak\'{o}w, Poland (e-mail: skowron@agh.edu.pl) }
%\thanks{K. Grochot and T. Stobiecki are also with AGH University of Science and Technology, Faculty of Physics and Applied Computer Science, Al. Mickiewicza 30, 30-059 Krak\'{o}w, Poland}
%\thanks{J. Langer, B. Ocker and M. Vafaee are with Singulus Technologies AG, Hanauer Landstrasse 103, Kahl am Main, 63796, Germany (e-mail: mehran.khanjani@singulus.de) }

\maketitle

%\date{\today}

\begin{abstract}
Materials with significant spin-orbit coupling enable efficient spin-to-charge interconversion, which can be utilized in novel spin electronic devices. A number of elements, mainly heavy-metals (HM) have been identified to produce a sizable spin current ($j_\mathrm{s}$), while supplied with a charge current ($j$), detected mainly in the neighbouring ferromagnetic (FM) layer. Apart from the spin Hall angle $\theta_\mathrm{SH}$ =  $j_\mathrm{s}$/$j$, spin Hall conductivity ($\sigma_\mathrm{SH}$) is an important parameter, which takes also the resistivity of the material into account. In this work, we present a measurement protocol of the HM/FM bilayers, which enables for a precise $\sigma_\mathrm{SH}$ determination. Static transport measurements, including resistivity and magnetization measurements are accompanied by the angular harmonic Hall voltage analysis in a dedicated low-noise rotating probe station. Dynamic characterization includes effective magnetization and magnetization damping measurement, which enable HM/FM interface absorption calculation. We validate the measurement protocol in Pt and Pt-Ti underlayers in contact with FeCoB and present the $\sigma_\mathrm{SH}$ of up to 3.3$\times$10$^5$ S/m, which exceeds the values typically measured in other HM, such as W or Ta.
\end{abstract}
    

\section{Introduction}
\IEEEPARstart{U}{tilizing} spin of the electron in addition to its charge opens up the way for the design of novel electronic devices\cite{hirohata2020}. The magnetic material, whose properties can be controlled using the spin current, is utilized widely in magnetic field sensors, data storage devices \cite{andrawis2018}, radio-frequency (rf) electronics \cite{Dieny2020} and, more recently, in hardware implementation of biomimetic circuits\cite{grollier2016}. It has been recently proposed, that non-magnetic elements with high spin-orbit coupling can act as a source of the spin current \cite{hirsch1999}. In such spin-orbit torque (SOT) effect, the spin current ($j_\mathrm{s}$) is generated as result of the charge current ($j$), with electron current, spin current and spin vectors being orthogonal to each other \cite{hoffmann2013}. Since it's discovery, there have been multiple studies on SOT in different materials exhibiting spin-orbit coupling and the effect is now quantified using so-called spin Hall angle ($\theta_\mathrm{SH}$) = $j_\mathrm{s}$/$j$ \cite{miron2011,liu2012}. Several methods of determination of $\theta_\mathrm{SH}$ were proposed, such as ferromagnetic resonance line-shape analysis\cite{liu2011}, harmonic Hall voltage measurement\cite{kim2013}, Kerr-effect-based optical determination\cite{stamm2017}, spin Hall magnetoresistance\cite{nakayama2013} and threshold current magnetization switching\cite{hao2015}. All these methods require slightly different mutlilayer structure, or different  ferromagnetic detecting layer's anisotropy axis and post-analysis, which enables quantitative investigation of material parameters. 

In this Letter, we present an experimental protocol for detailed measurement of spin Hall efficiency based on static and dynamic electrical measurements. We select conventional system composed of a Pt/FeCoB bilayer together with a recently proposed\cite{zhu2019} Pt-Ti/FeCoB system for the protocol verification. 
The presented protocol does not require special conditions, as opposed to perpendicular magnetic anisotropy of the FM for in-plane harmonic Hall voltage determination or negligible field-like torque for ferromagnetic resonance line-shape analysis.  
Pt was chosen as one of the most reliable material with well-established parameters, especially in terms of $\theta_\mathrm{SH}$, crystal structure and conductivity \cite{HweeWong2021}, while FeCoB has been shown to exhibit the highest tunneling magnetoresistance effect \cite{ikeda2008} together with the MgO barrier \cite{yuasa_giant_2004}. It has been recently shown, that creating additional interfaces\cite{zhu_prl2019}, for example by using Pt-Ti multilayer, $\theta_\mathrm{SH}$ can be enhanced at the cost of the resistivity. 
Utilizing a simple two-step lithography procedure, the Hall-bar matrix of different thickness of Pt or Pt-Ti interface number was fabricated, which enables determination of the magnetization saturation, resistivity, magnetization damping and spin Hall efficiency. Magnetotransport properties measurement protocol include both high-frequency dynamic characterization and low-frequency resistance and harmonic Hall voltage measurements characterized by an ultra low-noise. Investigation within the two frequency regimes allows us to determine all mentioned magnetotransport properties in a single device. For this a multi-probe radio-frequency rotating probe station was designed, which is controlled by a linear driver, enabling detailed measurement with a nV resolution. Using the presented protocol we investigate both Pt/FeCoB and Pt-Ti/FeCoB bilayer properties that are the most relevant for device applications.

\section{Experiment}
	\subsection{Deposition and microlithography}
Multilayer stacks were deposited using magnetron sputtering on thermally oxidized 4-inch Si wafer, with a wedged-shaped Pt layer in a Singulus Timaris cluster tool system. The following structure with Pt:
Ta(1)/Pt($t_\mathrm{Pt}$)/Fe$_{60}$Co$_{20}$B$_{20}$(2)/Ta(2) (thickness in nm) with $t_\mathrm{Pt}$ varying from 4 to 16 nm and Pt-Ti: Ta(1)/[Pt(\textit{d})/Ti(0.2)]$_m$/Pt(\textit{d})/Fe$_{60}$Co$_{20}$B$_{20}$(2)/Ta(2) where (\textit{m}+1)$\times$\textit{d} = 6 nm for m = 3, 5, and 7 interlayers, were deposited and annealed at 310$^\circ$C in ultra high vacuum chamber. The annealing step is required for a crystallization of the CoFeB along with the MgO tunnel barrier of the magnetic tunnel junction for further applications.  The slope of the Pt-wedge is 1.2 nm/cm. The top Ta layer oxidizes after exposing to the atmospheric condition, forming a protective layer, while the bottom Ta layer serves as a buffer. After the deposition process, the multilayers were patterned into 200 $\mu$m $\times$ 30 $\mu$m stripes for resistance measurements and 30 $\mu$m $\times$ 10 $\mu$m Hall-bars for static and dynamics transport characterization, using optical lithography, lift-off and ion-beam etching processes. The long axis of the stripes is perpendicular to the wedge direction, resulting in a negligible variation of the Pt thickness in a single device. Ti(5)/Au(50) contact electrodes were fabricated in a second lithography step. Devices enable resistance measurement using four-point method, harmonic Hall voltage measurements and ferromagnetic resonance (FMR) using so-called spin-diode effect \cite{Tulapurkar2005}.

\begin{figure}[t]
\centering
\includegraphics[width=\columnwidth]{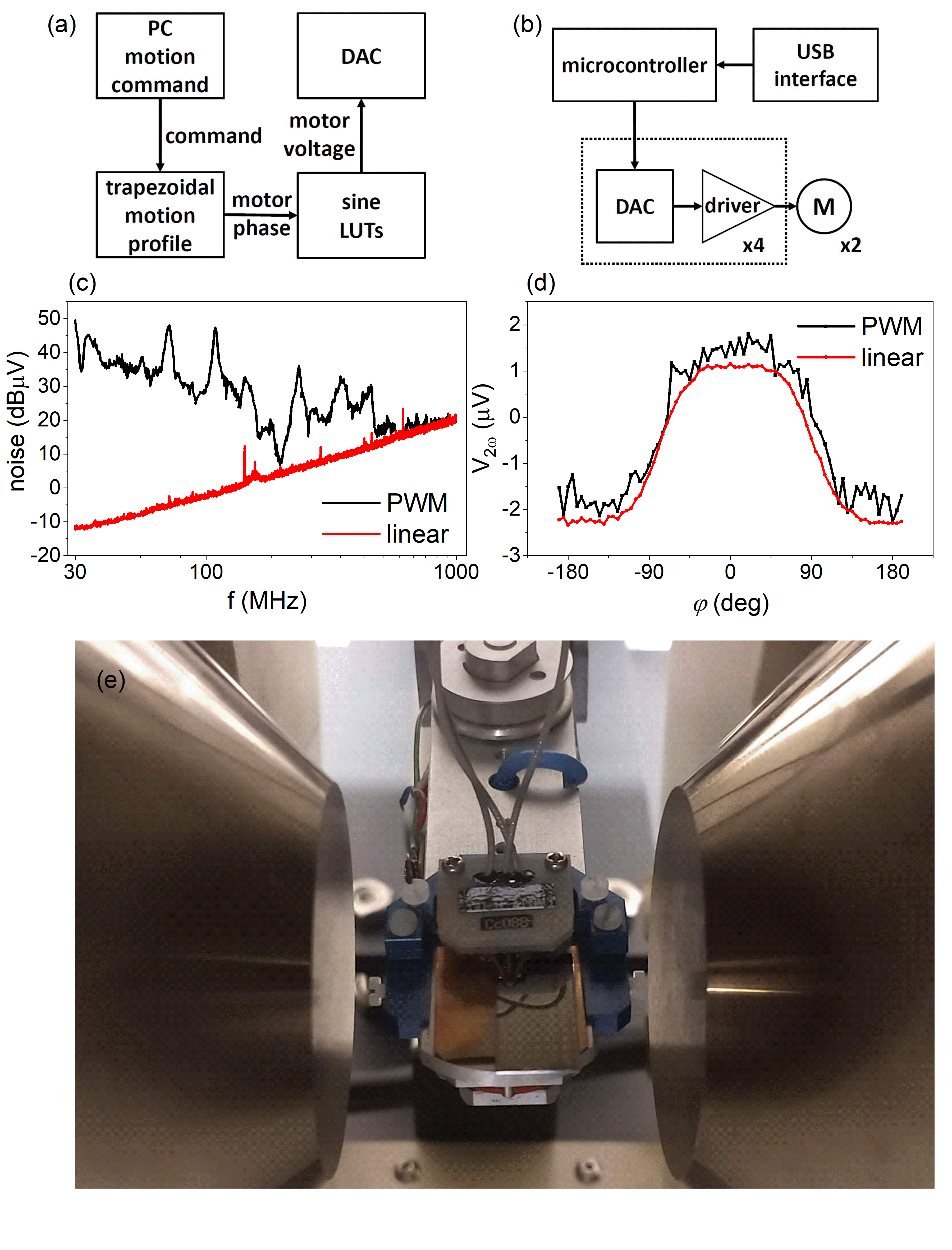}
\caption{Microwave rotating probe station design: software flow (a), driver's hardware block-diagram (b), comparison of the switching and linear driver performance in terms of the electromagnetic radiation (c) and second harmonic Hall measurements (d). A photograph of the rotating probe station is presented in (e).}
\label{fig:fig1}
\end{figure}

	\subsection{Transport measurement}
The Hall voltage signal was measured using low-frequency (2387 Hz) stimuli of 1 V in both out-of-plane and in-plane magnetic field. Depending on the sample resistance, the current (current density) ranged from 2.2 to 10 mA (5.2 to 7.2 MA/cm$^2$). FMR was measured using amplitude-modulated radio-frequency (RF) signal of $P$ = 16 dBm applied to the long axis of the stripe and measurement of the mixing voltage ($V_\mathrm{mix}$) using short-axis electrodes with a lock-in amplifier synchronized with the modulating signal. Harmonic Hall voltage measurements were performed in a fixed magnetic field with a rotating sample stage with an rf-mulitprobe from T-plus attached. The mechanical rotation of the stage was controlled with a stepper motor driven by the dedicated linear-controller.
	\subsection{Rotating probe station}
Detailed investigation of the spin Hall effect requires measurement in an magnetic field applied at various angles with respect to the Hall-bar axis. Specifically, anisotropy field $H_\mathrm{k}$ is determined from anomalous Hall effect (AHE) measured in a perpendicular magnetic field, while harmonic Hall voltage investigation requires rotation of the in-plane magnetic field vector. Dedicated mechanical setup was designed and fabricated by Measline Ltd., which enabled rotation of the sample at an arbitrary polar and azimuth angle with respect to the magnetic field produced by the dipole electromagnet. To reduce the noise, a dedicated stepper driver controller was designed.

A standard stepper motor driver incorporates pulse width modulation (PWM) signal to tune and stabilize coil current of the motors. This current must be applied to the motor even when stationary, to hold the position while using micro-stepping (i.e. ability to move motor more precise than its full step resolution). While PWM signal allows to reduce size of the driver unit and increase it's power efficiency, it also generates a switching noise. This causes generation of dynamic electromagnetic fields around the motors and cabling, which significantly disrupts precise measurements performed in the rotating setup due to Radio Frequency Interference (RFI). The spectra of this interference are broad, as it originates from a square-shape PWM waveform. In order to eliminate it, a specialized driver was designed, which uses a  linear (i.e. non-switching) voltage sources for the motors with negligible RFI. Additionally, a dedicated printed circuit board was designed to minimize the Electromagnetic Interference from the board itself.

The stepper motor driver was built using multiple high-current digitally controlled voltage sources. Instead of using a PWM signal, Digital to Analog Converters (DAC) - dual-channel 14-bit AD5643R - provide a voltage corresponding to the requested motor's phase current. Operational amplifiers - OPA548T - uses 0-2.5 V voltage from DAC and convert it to bipolar $\pm$9 V. Each coil of the motor has its own DAC and the driver, in addition to the total of four drivers - Fig. \ref{fig:fig1}(b). An 32-bit ARM microcontroller STM32F103C8T6 is responsible for receiving commands from the measurement software using FT230XQ USB interface, creating trapezoidal motion profiles (to smoothly start and end the movement) and controlling DACs by generating required voltages for each motor phase. Maximum voltages per phase, parameters of trapezoidal motion profiles and microstepping factors can be adjusted by sending configuration commands. Whole system is powered using external linear laboratory power supply, which is also controlled by a designed unit. Additionally, the USB interface is completely opto-isolated from the system, which protects the PC in case of mains power issues. A software flow is shown in Fig. \ref{fig:fig1}(a).

\begin{figure}[t]
\centering
\includegraphics[width=\columnwidth]{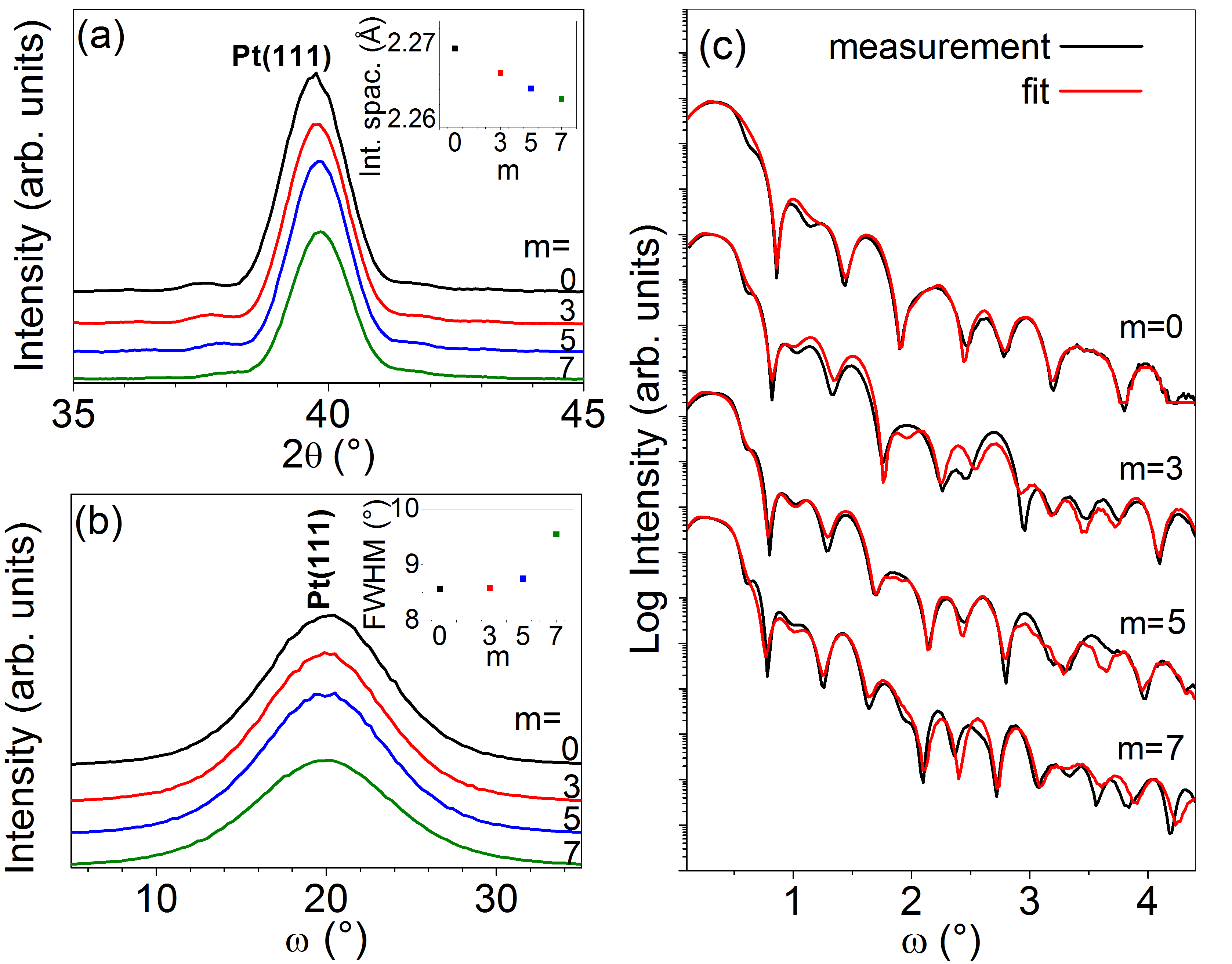}
\caption{X-ray diffraction and X-ray reflectivity measurements of the investigated multilayers. Intensity of the Pt(111) peak decreases with increasing Ti interfaces (a). In addition, the peak position shifts to higher angles with higher \textit{m}, which is interpreted as a decrease in interplanar distances of Pt (inset). Rocking curves are presented in (b) resulting in an increase of FWHM with \textit{m} (inset). XRR for all multilayers together with corresponding fits are shown in (c). The fitting parameters are collected in Table \ref{tab:table1}.}
\label{fig:fig1b}
\end{figure}

Figure \ref{fig:fig1}(c) presents the comparison of the measured radiation emissions correlated to 10 m open area test site (OATS) for both PWM (based on Leadshine EM402 driver) and the presented linear driver. Measurements were performed using GHz transverse electromagnetic (GTEM) chamber, calculated and correlated according to electromagnetic interference standards \cite{guirado1995}. To verify the operation of the driver the angular measurement of the second harmonics Hall voltage dependence on the in-plane magnetic field was determined - Fig. \ref{fig:fig1}(d) - which shows an improvement with respect to the PWM-based driver in terms of the measurement noise. An image of the entire probe station is presented in Fig. \ref{fig:fig1}(e). Specifically, both in-plane and out-of-plane magnetic field characteristics as well as angular dependence using 4-point technique can be measured upon contacting the device under test, which is an advancement to the previously published setup \cite{he2016, Tamaru_2014}.

\section{Results and discussion}
	\subsection{X-ray  diffraction, resistivity and anomalous Hall effect}
X-ray diffraction (XRD) patterns for both Pt- and Pt-Ti-based multilayers are presented in Fig. \ref{fig:fig1b}(a). The intensities of $\theta$-2$\theta$ XRD peaks decrease together with increasing the number \textit{m} of the Pt-Ti multilayers. The lattice plane spacing calculated from the position of Pt (111) decreases similarly to \cite{zhu2019} - Fig. \ref{fig:fig1b}(a) inset. For both Pt and Pt-Ti based samples, the grain size in direction perpendicular to the layers, calculated from the Scherrer equation, are comparable to the Pt and Pt-Ti thickness. Figure \ref{fig:fig1b}(b) shows XRD rocking curve patterns of Pt(111) for investigated structures. From fitting of the rocking curve peaks a Full Width at Half Maximum (FWHM) parameter was determined (Figure \ref{fig:fig1b}(b) inset). The highest FWHM was calculated for the sample with the greatest number of Ti layers, which means that  the Ti atoms diffused into Pt grains \cite{Kopcewicz_1997} after the deposition and annealing process.
X-ray reflectivity (XRR) measurements and the corresponding fits for the Pt-Ti based multilayers are shown in Fig. \ref{fig:fig1b}(c). Fitting parameters for samples with different number of Ti monolayers are collected in Table \ref{tab:table1}. The fits were made by replacing the Pt-Ti superlattice with a single Pt layer with total [Pt(\textit{d})/Ti(0.2)]$_m$/Pt(\textit{d}) thickness and reduced density (compared to the Pt density of 21.45 g/cm$^3$) as a result of dissolving Ti in Pt. This assumption was supported by a strong mixing enthalpy for Ti in Pt (-290 kJ/mole of atoms) and Pt in Ti (-290 kJ/mole of atoms) \cite{Boer1988}.

%\centering
\begin{table*}
\caption{XRR fitting parameters - density (g/cm$^3$) , thickness (nm), roughness (nm) - for Pt and Pt-Ti-based multilayers.}
\label{tab:table1}
\setlength{\tabcolsep}{7pt}
%\begin{ruledtabular}
%\textwidth
\begin{tabular}{| l | c c c | c c c | c c c | c c c |}
\hline
  &  & Pt & & & [Pt-Ti]$_3$ & & & [Pt-Ti]$_5$ & & & [Pt-Ti]$_7$ & \\
 & density & \textit{t} & roughness &  density & \textit{t} & roughness & density & \textit{t} & roughness  & density & \textit{t} & roughness  \\
 \hline
Si & 2.33 & -& 0.27& 2.33 &	- & 0.26 &	2.32 & - &	0.31 & 2.33 & - & 0.26 \\
SiO & 2.64 &100& 0.31& 2.64	&100&	0.29&	2.64&	100&	0.34&	2.64&	100&	0.26\\
Ta	& 16.6 &1.10& 0.46&13.3&	0.86&	0.24&	14.1&	0.99&	0.21&	13.6&	0.98&	0.35\\
Pt	& 21.4 &5.97& 0.47&16.9&	6.69&	0.44&	17.4&	7.05&	0.48&	17.0&	7.44&	0.43\\
FeCoB & 8.13 &1.89& 0.36&8.31&	1.88&	0.42&	8.29&	1.87&	0.37&	8.38&	1.76&	0.44\\
Ta	& 15.9 &0.36& 0.50&13.3& 1.36&	0.36&	13.5&	1.26&	0.34&	13.0&	1.35&	0.36\\
Ta$_2$O$_5$ & 5.14 & 3.46& 0.37& 5.86&	3.27&	0.36&	6.64&	3.49&	0.41&	5.57&	3.31&	0.37\\

\hline
\end{tabular}
%\end{ruledtabular}
\end{table*}

\begin{figure}[t]
\centering
\includegraphics[width=\columnwidth]{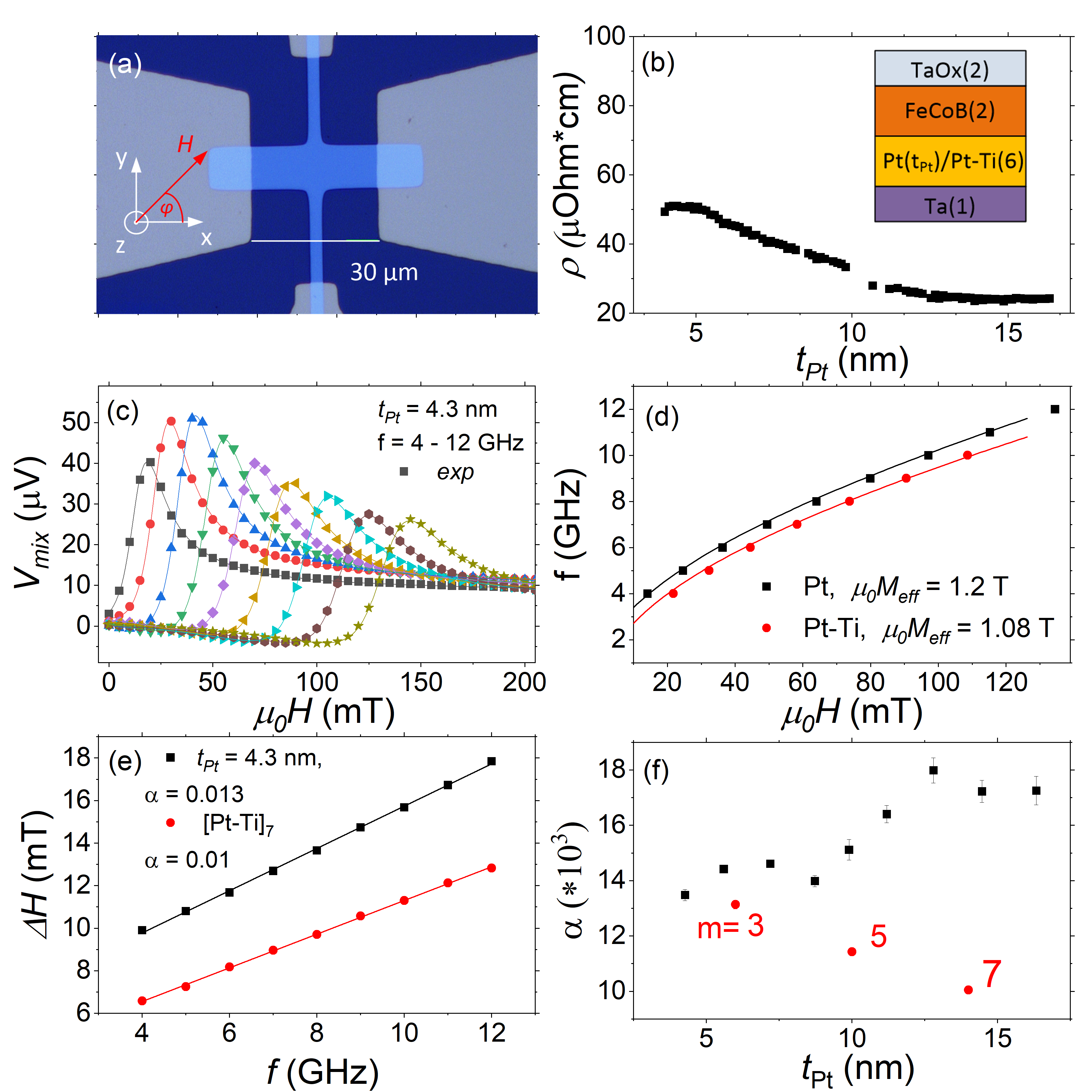}
\caption{Micrograph of the Hall-bar (a), resitivity vs. Pt thickness dependence (b), example of the FMR measured using spin-diode effect for different excitation frequency for the Hall-bar with $t_\mathrm{Pt}$ = 4.3 nm (c), resonance frequency vs. in-plane magnetic field for Pt ($t_\mathrm{Pt}$ = 4.3 nm) and Pt-Ti (m = 7) underlayers modelled using Kittel formula (d), FMR linewidth ($\Delta H$) vs. frequency dependence, together with a line fit, enabling damping constant calculation (e) and damping dependence on $t_\mathrm{Pt}$ and number of Pt-Ti interfaces (f).}
\label{fig:fig2}
\end{figure}

Around one hundred devices with different $t_\mathrm{Pt}$ were used for resistivity ($\rho$) determination. For each device, the resistance ($R$) was transformed into sheet conductance $G_\mathrm{S}$ = $l$/$wR$, with $l$ and $w$ being the length and the width of the stripe. The resistivity of Pt was than calculated as $\rho_\mathrm{Pt}$ =  $t_\mathrm{Pt}$/($G_\mathrm{S}$-$G_\mathrm{0}$), where $G_\mathrm{0}$ = 0.0011 S is the sheet conductance of the multilayer without Pt buffer. Similarly, the resistivity of the Pt-Ti $\rho_\mathrm{Pt-Ti}$ was calculated for different number of Pt-Ti interfaces. The results are presented in Fig. \ref{fig:fig2}, together with a micrograph of the Hall-bar. $\rho_\mathrm{Pt}$ vs. $t_\mathrm{Pt}$ behaviour resembles similar dependence measured in Pt/Co/MgO multilayers\cite{nguyen2016}. %The values of $\rho_\mathrm{Pt-Ti}$ agrees with Ref. \cite{zhu2019}. In addition to resistivity measurements, one also need a anisotropy field $H_\mathrm{k}$, which was determined from the anomalous Hall effect measurement in the perpendicular magnetic field. 

	\subsection{Saturation magnetization and damping}
Saturation magnetization ($M_\mathrm{S}$) for each underlayer was measured using vibrating sample magnetometer, while the effective magnetization ($M_\mathrm{eff}$) and magnetization damping ($\alpha$), were determined from the FMR. To do so, the $V_\mathrm{mix}$ vs. in-plane magnetic field ($H$) was measured for different input frequencies ($f$) spanning from 4 to 12 GHz. As a result, the family of Lorentz-shape curves were obtained, which were fitted using the formula:
 \begin{equation}
V_{\mathrm{mix}} = V_\mathrm{S}\frac{\Delta H^2}{\Delta H^2+(H-H_0)^2} + V_\mathrm{AS} \frac{\Delta H(H-H_{0})}{\Delta H^2+(H-H_{0})^2}\,
\label{eq:Vmix}
\end{equation}
where $V_\mathrm{S}$ and $V_\mathrm{AS}$ are the magnitude of symmetric and antisymmetric Lorentz curves, respectively, $\Delta H$ is the linewidth, and $H_\mathrm{0}$ is the resonance field. $f$ vs. $H_\mathrm{0}$ dependence was fitted using the Kittel formula \cite{kittel1948}, the fitting parameters are presented in Table \ref{tab:table2}. The dependence of damping on Pt thickness and Pt-Ti interfaces is depicted in Fig. \ref{fig:fig2}(f). We note, that for Pt-Ti buffer, the effective damping decreases with increasing number of interfaces (m), which may be useful for applications, as the critical current density needed to switch the ferromagnet depends linearly on the effective damping\cite{myers1999}.

	\subsection{Spin Hall angle}
\begin{figure}[t]
\centering
\includegraphics[width=\columnwidth]{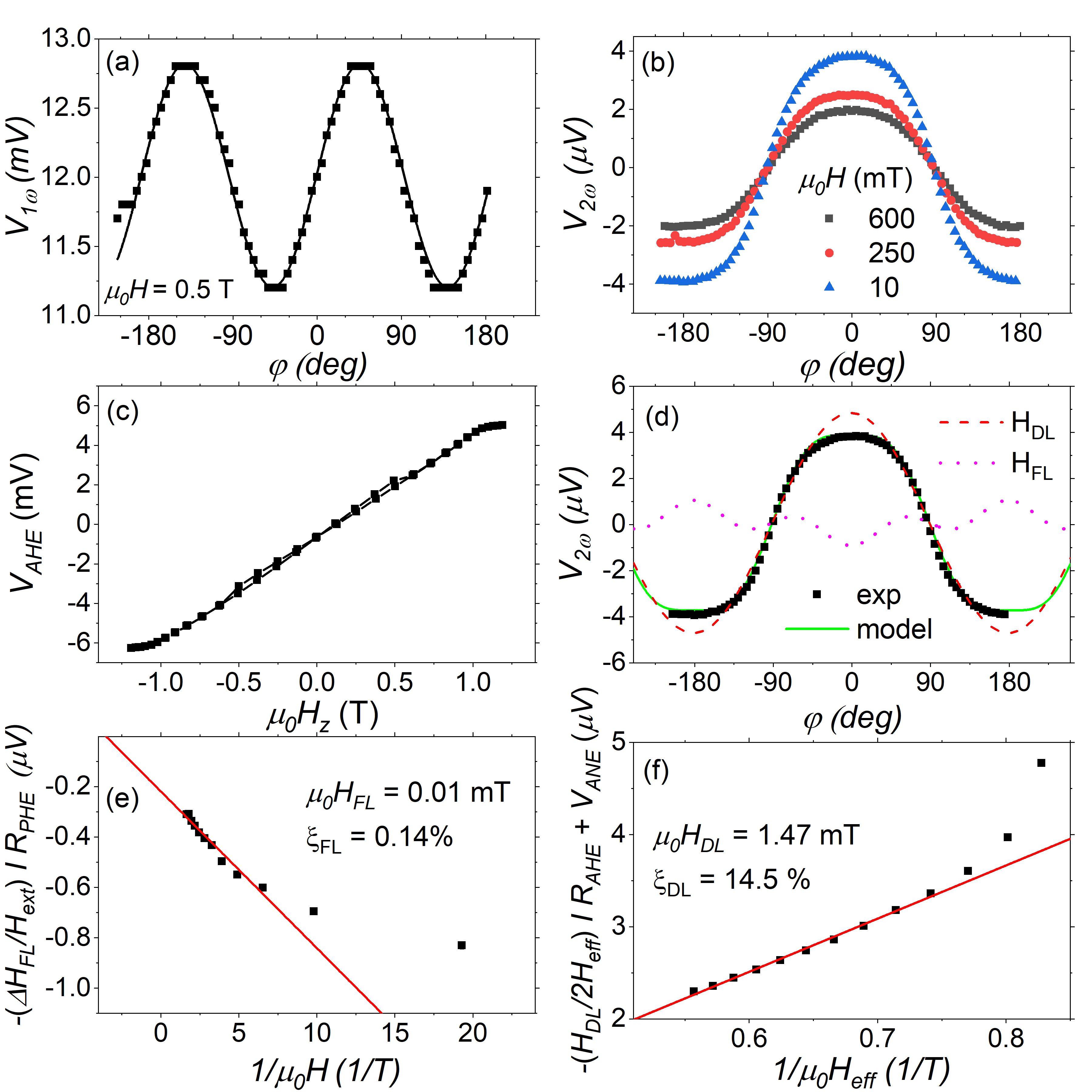}
\caption{Hall voltage measurements of the first (a) and second (b) harmonic component vs. in-plane magnetic field angle of the Hall bar with $t_\mathrm{Pt}$ = 4.3 nm. Anomalous Hall voltage measurement in the perpendicular field, i.e., along z axis enables anisotropy field determination (c). Second harmonic Hall voltage was modelled using Eq. \ref{eq:Vharm} enabling separation of the damping-like ($H_\mathrm{DL}$) and field-like ($H_\mathrm{FL}$) effective field components (d). Dependence of the $H_\mathrm{FL}$ vs. inverse of the applied magnetic field (e) and $H_\mathrm{DL}$ vs. inverse of the effective magnetic field (f) enables calculation of the spin Hall efficiencies.}
\label{fig:fig3}
\end{figure}

The spin Hall efficiency was determined from the harmonic Hall voltage measurements performed in the rotating in-plane magnetic field\cite{avci2014}. Figure \ref{fig:fig3} presents angular dependence of the first ($V_\mathrm{\omega}$) and second ($V_\mathrm{2\omega}$) harmonic Hall voltage vs. azimuth angle $\varphi$ between magnetic field vector and long axis of the Hall-bar (as denoted in Fig.\ref{fig:fig3}(a)). Both dependencies were modelled using Eq:
 \begin{equation}
 \begin{split}
V_{\mathrm{\omega}} = V_\mathrm{P} \sin2\varphi \\
V_{\mathrm{2\omega}} = -\frac{V_\mathrm{P}\left(H_\mathrm{FL}+H_\mathrm{Oe}\right)}{H}\cos\varphi \cos2\varphi + \\
- \left(\frac{V_\mathrm{A}H_\mathrm{DL}}{2 H_\mathrm{eff}} - V_\mathrm{ANE}\right)\cos\varphi
\end{split}
\label{eq:Vharm}
\end{equation}
where $V_\mathrm{P}$ and $V_\mathrm{A}$ are the planar and anomalous Hall voltages, $H_\mathrm{FL}$, $H_\mathrm{Oe}$ and $H_\mathrm{DL}$ are the field-like, Oersted field and damping-like field components, $H_\mathrm{eff}$ = $H$ + $H_\mathrm{k}$ is the effective field, which is the sum of the external magnetic field and anisotropy field, determined from the AHE measurement in the perpendicular orientation ($H$ along $z$ axis). $V_\mathrm{ANE}$ is the additional contribution from the anomalous Nernst effect, which was found negligible in the investigated bilayers. 
The spin Hall efficiency $\xi_\mathrm{DL}$ was calculated using the formula:
 \begin{equation}
\xi_\mathrm{DL} = \cfrac{\mu_\mathrm{0} M_\mathrm{S} t_\mathrm{FeCoB} H_\mathrm{DL}}{j \hbar/2e}
\label{eq:ksi}
\end{equation}

As a result of the calculation, the dependence of the spin Hall efficiencies were obtained as a function of $t_\mathrm{Pt}$. Similar to Ref.\cite{nguyen2016, pai2015} $\xi_\mathrm{DL}$ reaches its maximum for $t_\mathrm{Pt}$ = 5 nm, and decreases with increasing thickness of HM. To account for the Oersted field we used an analytical expression $H_\mathrm{Oe}$ = $J t_\mathrm{Pt}$/2. The calculated values dominate the field-like contribution, which means that both $H_\mathrm{FL}$ and $H_\mathrm{Oe}$ are comparable in amplitude but of opposite sign. Nonetheless, $\xi_\mathrm{FL}$ is an order of magnitude smaller than $\xi_\mathrm{DL}$ for the thinnest Pt but increases with increasing $t_\mathrm{Pt}$. The results of the effective field values are presented in Fig. \ref{fig:fig4} 
%for $t_\mathrm{Pt}$ = 5 nm and increases to around 2 \% for maximum Pt thickness investigated, which agrees with the reports on in-plane magnetized ferromagnets\cite{pai2015}. The detailed data are included in Table \ref{tab:table1}.
In order to calculate the intrinsic $\theta_\mathrm{SH}$ one has to take into account the interface transparency. To do so, we first calculate the absorption of the spin current on the Pt/FeCoB interface using the following formula:
\begin{equation}
g^{\updownarrows}_\mathrm{eff} = \frac{4 \pi M_\mathrm{S} t_\mathrm{FeCoB}}{g \mu_\mathrm{B}}(\alpha_\mathrm{eff}-\alpha_0)
\label{eq:absorption}
\end{equation}
where $g$ is the Lande factor, $\mu_\mathrm{B}$ is the Bohr magneton and $\alpha_0$ = 0.004 is intrinsic damping of the FeCoB. $g^{\updownarrows}_\mathrm{eff}$ for different Pt thickness and Pt-Ti underlayer were calculated and presented in Fig. \ref{fig:fig4}. In general $g^{\updownarrows}_\mathrm{eff}$ increases (decreases) with the thickness of Pt (with a number of Pt-Ti interfaces) as a result of increasing (decreasing) damping constant. Next, the interface transparency $T$ is calculated as:
 \begin{equation}
T=\cfrac{g^{\updownarrows}_\mathrm{eff} \mathrm{tanh}\cfrac{t_\mathrm{Pt}}{2 \lambda_\mathrm{Pt}}}{g^{\updownarrows}_\mathrm{eff} \mathrm{coth}\cfrac{t_\mathrm{Pt}}{\lambda_\mathrm{Pt}}+\cfrac{h}{\rho_\mathrm{Pt} \lambda_\mathrm{Pt} 2 e^2}}
\label{eq:transparency}
\end{equation}
where, $\lambda_\mathrm{Pt}$ = 2 nm is the spin diffusion length \cite{magni2021} in Pt, $h$ is the Planck constant and $e$ is the electron charge. The calculated transparency, which varies slightly between 0.42 and 0.5 for Pt/FeCoB and Pt-Ti/FeCoB interfaces were included in Table \ref{tab:table2}. Finally, $\theta_\mathrm{SH}$ = $\xi_\mathrm{DL}$/$T$ and spin Hall conductivity  $\sigma_\mathrm{SH}$ = $\xi_\mathrm{SH}$/$\rho_\mathrm{HM}$ were calculated. We note that we took into account the transparency of the HM/FM interface to calculate the intrinsic spin Hall angle \cite{HweeWong2021}. The spin Hall efficiency ranges between 0.17 and 0.05 depending on the thickness of Pt. Similar values were obtained using a line-shape analysis of the SOT-FMR signal, based on Ref. \cite{MacNeill2017}, SMR and the current-assisted magnetic-field switching experiments \cite{magni2021}. Nevertheless, the intrinsic spin Hall angle is a function of the interface transparency determination, which depends on spin diffusion length and spin mixing conductance. In the presented analysis, the interface transparency ranges between 0.5 and 0.4, which results in $\theta_\mathrm{SH}$ between 0.35 and 0.12. Smaller interface transparency would translate into even higher intrinsic $\theta_\mathrm{SH}$, which were recently reported in Refs. \cite{zhu2019,Li_PhysRevMaterials_2021}.

	\subsection{Magnetoresistance}
\begin{figure}[t]
\centering
\includegraphics[width=\columnwidth]{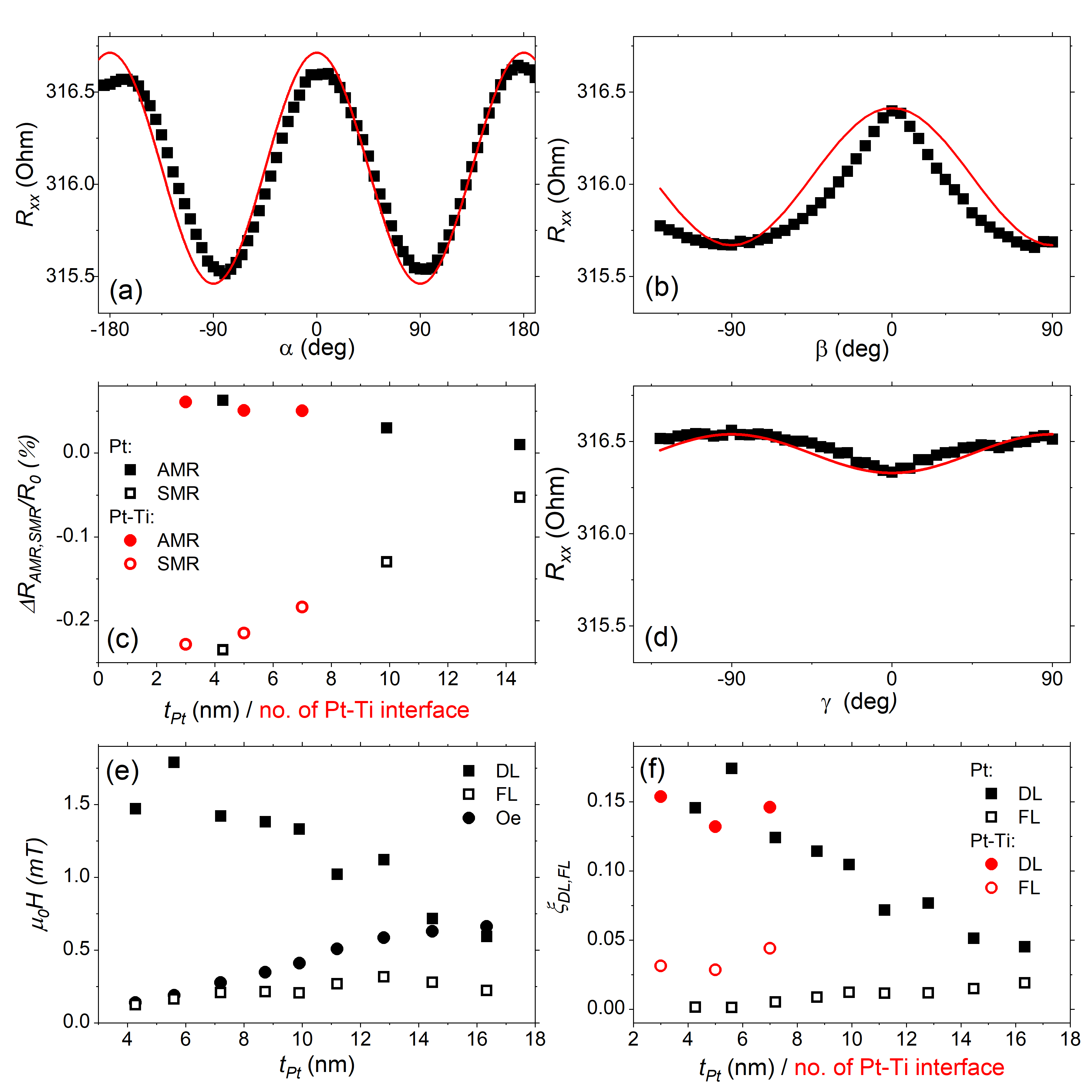}
\caption{Example of the longitudinal resistance ($R_\mathrm{xx}$) in different azimuth ($\alpha$) and polar ($\beta$ and $\gamma$) angles (a-c) for $t_\mathrm{Pt}$ = 4.3 nm, enabling separation of the AMR and SMR contribution to the total magnetoresistance. Measurements were performed in a magnetic field $H$ = 1.8 T, which exceeds the anisotropy field, but is unable to saturate the FeCoB magnetization at polar angles 0 < $\theta$ < 90, hence a small deviation from the cosine dependence in (b) and (d). AMR and SMR values determined for different $t_\mathrm{Pt}$ and no. of Pt-Ti interfaces are presented in (d). Calculated effective fields in Pt-based elements are presented in (e). As a result, the spin Hall efficiencies for all devices are shown in (f). }
\label{fig:fig4}
\end{figure}

To complete the spin-dependent transport analysis, the angular measurements of the magnetoresistance in x-y ($\alpha$-rotation), y-z ($\beta$-rotation) and x-z ($\gamma$-rotation) planes were conducted \cite{Cho2015}. Figure \ref{fig:fig4} presents angular resistance dependence in Pt/FeCoB with $t_\mathrm{Pt}$ = 4.3 nm. Using the formulas from Ref. \cite{kim2016} spin Hall magnetoresistance (SMR) and anisotropic magnetoresistance (AMR) were derived - Fig. \ref{fig:fig4}(c). The dependence of SMR on $t_\mathrm{Pt}$ and a number of Pt-Ti interfaces coincides with the dependence of the spin Hall conductivity, however, it does not match the dependence of the resistivity - for Pt underlayer the highest SMR was determined in the bilayer with the thinnest Pt, whereas for Pt-Ti the highest SMR was reported for the smallest number of Pt-Ti interfaces. The AMR contribution decreases with increasing $t_\mathrm{Pt}$ due to shunting effect from  Pt \cite{Karwacki2020}.

\begin{table}
\caption{
\label{tab:table2}
Summary of the spin transport parameters determined in Pt-based (different thickness) and Pt-Ti-based (different number of interfaces) devices}
%\begin{ruledtabular}
\begin{tabular}{ | l | c c c | c c c | c | }
\hline
 $t_{\text{Pt}}$/ &  & Pt & & & Pt-Ti & & units \\

Pt-Ti intf. &  5.6 & 10 & 14 & 3 & 5 & 7 & nm/no. \\
 \hline
$\rho_{\text{HM}}$ & 52 & 34 & 24 & 54 & 60 & 95 & $\mu\Omega\text{cm}$\\
$\mu_0M_{\textrm{S}}$ & 1.51 & 1.42 & 1.4 & 1.38 & 1.38 & 1.38 & $\text{T}$ \\
$\alpha$ & 14 & 15 & 17 & 13 & 11 & 10 & 10$^{-3}$\\
$g^{\updownarrows}_\mathrm{eff}$ & 16.3 & 16.8 & 19.6 & 13.4 & 10.4 & 8.9 &  10$^{18}$/m$^2$ \\
%$\xi_\mathrm{DL}$ & 17.4 & 10.4 & 5.1 & 17.6 & 15.2 & 16.6 & \% \\
%$\xi_\mathrm{FL}$ & 0.1 & 1.4 & 1.6 & 4.9 & 4.7 & 6 & \% \\
$T$& 0.5 & 0.46 & 0.42 & 0.48 & 0.44 & 0.51 & \\
$\theta_\mathrm{SH}$ & 0.35 & 0.23 & 0.12 & 0.36 & 0.34 & 0.32 & \\
$\sigma_\mathrm{SH}$ & 3.3 & 3.07 & 2.13 & 3.2 & 2.6 & 1.75 &  10$^5$/($\Omega m$)\\
\hline
\end{tabular}
%\end{ruledtabular}
\end{table}

There are a few materials reported with higher intrinsic spin Hall angle, including HMs: W \cite{pai2012, skowronski2016}, Ta \cite{liu2011} and topological insulators: BiSe \cite{DC2018}; however, at the cost of increased resistivity of the material. Similar tendency is maintained by alloying HMs with good conductors: AuPt \cite{zhu2020}, AuTa\cite{laczkowski2016}. Surprisingly,  $\sigma_\mathrm{SH}$ for Pt reaches a maximum value of 3.3$\times$10$^{5}$ S/m for 5.6 nm-thick underlayer, which is among one of the highest values reported for Pt \cite{Savero_Torres_2017,Yan2017, Pham2016}, Pt-based  multilayers \cite{zhu_prl2019} and alloys \cite{zhu2018}, exceeding other HM and their compounds \cite{kim2020}, mainly due to small resistivity of the Pt. Even higher intrinsic spin Hall angle has been measured recently in an all-epitaxial ferrite/Pt system \cite{Li_PhysRevMaterials_2021}. This findings, together with well-established deposition technique, crystal properties, and endurance leads to the conclusion that Pt is one of the most optimal materials for spin-orbit torque applications in line with potential spin-logic circuits \cite{Pham2020}.

\section{Summary}
In summary, the measurement protocol of the spin orbit torque efficiency based on low-frequency harmonic Hall voltage measurement and radio-frequency ferromagnetic resonance analysis was presented. The design of the setup including a rotating probe station and a dedicated linear driver together with microfabricated Hall bar bilayers based on FeCoB with different underlayer materials was shown, which enables for thorough analysis of the spin-dependent phenomena in HM/FM bilayer system. The protocol was verified in Pt and Pt-Ti based devices, resulting in an intrinsic spin Hall angle of up to 0.35 and spin Hall conductivity reaching 3.3$\times$10$^{5}$ S/m, which is among the highest values reported. Our findings indicate that Pt remains one of the most attractive material for spintronics. 

%\section*{Acknowledgments}
%This work is supported by the National Science Centre, grant No. UMO-2015/17/D/ST3/00500, Poland. K.G. and S.Ł. acknowledge support from National Science Centre, grant No. UMO-2016/23/B/ST3/01430, Poland. Nanofabrication was performed at the Academic Centre for Materials and Nanotechnology of AGH University of Science and Technology.

%\nocite{*}
\bibliographystyle{ieeetr}
\bibliography{bibliography}% Produces the bibliography via BibTeX.

\end{document}